\documentclass[a4paper]{jpconf}
\usepackage{graphicx}
\begin{document}
\title{Anisotropic fragmentation in low-energy dissociative recombination}

\author{S.~Novotny$^1$, H.~Rubinstein$^2$, H.~Buhr$^{1,2}$, O.~Novotn\'y$^1$, J.~Hoffmann$^1$, M.B.~Mendes$^1$, D.A.~Orlov$^1$, M.H.~Berg$^1$, M.~Froese$^1$, A.S.~Jaroshevich$^{1,3}$, B.~Jordon-Thaden$^1$, C.~Krantz$^1$, M.~Lange$^1$, M.~Lestinsky$^1$, A.~Petrignani$^1$, I.F.~Schneider$^4$, D.~Shafir$^2$, F.O.~Waffeu Tamo$^4$, D.~Zajfman$^2$, D.~Schwalm$^{1,2}$ and A.~Wolf$^1$}
\address{$^1$Max-Planck-Institut f\"ur Kernphysik, D-69117
  Heidelberg, Germany}
\address{$^2$Weizmann Institute of Science, Rehovot 76100, Israel}%
\address{$^3$Institute of Semiconductor Physics, 630090 Novosibirsk, Russia}
\address{$^4$University of Le Havre, 76058 Le Havre, France}
 
\ead{steffen.novotny@mpi-hd.mpg.de}

\begin{abstract}
On a dense energy grid reaching up to 75 meV electron collision energy the fragmentation angle and the kinetic energy release of neutral dissociative recombination fragments have been studied in a twin merged beam experiment. The anisotropy described by Legendre polynomials and the extracted rotational state contributions were found to vary on a likewise narrow energy scale as the rotationally averaged rate coefficient. For the first time angular dependences higher than 2$^{nd}$ order could be deduced. Moreover, a slight anisotropy at zero collision energy was observed which is caused by the flattened velocity distribution of the electron beam.
\end{abstract}

\section{Introduction}

In low-energy dissociative recombination (DR) two pathways of compatible strength, the direct and the indirect mechanism~\cite{bardsley68}, compete with each other and lead to interference effects resulting in rich resonant structures of the DR cross section~\cite{schneider:97}. These occur as both of them can often access the same neutral doubly excited dissociative state, which is formed by purely electronic energy exchange in the direct process and through intermediate ro-vibrationally excited Rydberg states in the indirect process. Non-adiabatic coupling mechanisms between the electronic and nuclear motion in the indirect process further cause the DR reaction to become particularly sensitive on the ro-vibrational state of the molecular ion~\cite{takagi:93,schneider:97}.

On the experimental side the resonant energetic structures in the DR cross section have so far mainly been studied in event-by-event counting experiments which have revealed for the deuterated hydrogen molecular ion HD$^+$ pronounced narrow resonances down to meV electron collision energies~\cite{alkhalili02}. In the present experiment we have focused for the first time on the nature of the resonances appearing in HD$^+$ DR at energies below $\sim 75$ meV through measurements of the fragment kinetic energy release (KER) and the angular distribution. Thereby the fragment angular distribution can be related to the angular dependence of the electron capture reaction, i.e. its dependence on the angle relative to the molecular axis. However, the fragment direction reflects the initial orientation of the molecule only if its rotation proceeds slowly in comparison to the dissociation time; only in this case the axial recoil approximation~\cite{zare:67,malley:68,guberman:04} is justified. 

The measurements presented here have been made possible by the new twin-merged beam setup with nearly monochromatic electrons at the heavy ion storage ring TSR in Heidelberg, Germany. Fragment imaging experiments could thus be performed under stable ion beam conditions and variable low electron collision energies. The use of a photocathode electron source~\cite{orlov04} provided an electron beam with low energy spread for high precision measurements down to the meV range. These conditions enabled us to obtain both the KER as well as the product angular distribution on a dense energy grid. The rotational levels of the molecular ions contributing to the resonances could then be extracted from the KER and revealed similar as the measured angular distribution likewise narrow variations as a function of energy as the DR rate coefficient.

The HD$^+$ molecular ion has been chosen for a number of theoretical and experimental studies and has become, due to its simple structure, a benchmark system in DR research. Moreover, the stored infrared active HD$^+$ ion cools down to the vibrational ground state within hundreds of ms~\cite{amitay98} and reaches an equilibrium rotational state distribution mainly determined by the surrounding room temperature after only a few seconds. At low electron collision energies $E_d$ the HD$^+$ DR process is then dominated by the doubly excited neutral HD potential curve $(2p\sigma_u)^2$ of $^1\Sigma^+_g$ symmetry which leads to only one final state configuration, \mbox{[H($n=$ 1)+D($n=$ 2)]} or \mbox{[H(2)+D(1)]}, for $E_d<1.14$~eV~\cite{zajfman:95}. This simplifies the analysis of the KER and the HD$^+$ ion thus becomes an ideal study case on the DR angular dependence at low collision energies.

\section{Experimental}
\subsection{Setup}
\begin{figure}[b]
\center
\includegraphics[width=9cm,keepaspectratio]{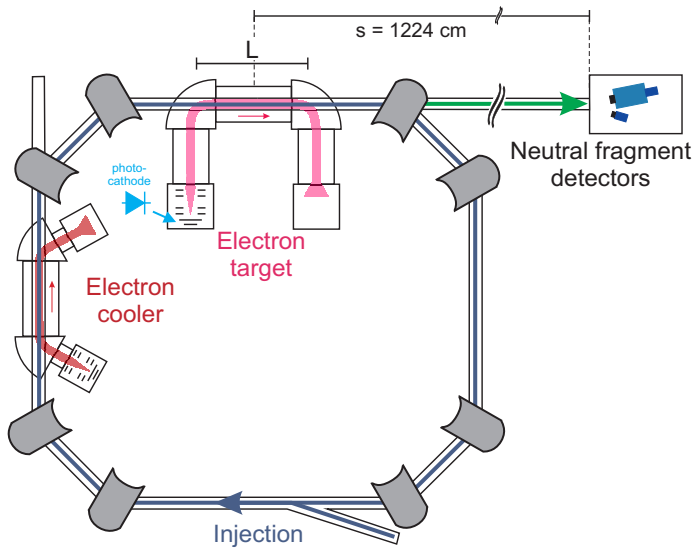}\hspace{1pc}%
\begin{minipage}[b]{6cm}\caption{\label{fig:tsrsetup} The twin-merged beam setup at the TSR consisting of the electron-cooler and electron-target section (ETS). Neutral fragments are recorded downstream of the ETS by counting or fragment imaging detectors.}
\end{minipage}
\end{figure} 
In the present experiment, an HD$^+$ ion beam is produced in a standard Penning ion source and accelerated to \mbox{1.44 MeV} by the HSI accelerator~\cite{grieser:93} before injection into the storage ring TSR~\cite{habs:89}. Compared to earlier experiments the ion beam is merged with electron beams in two separate sections of the storage ring (Fig.~\ref{fig:tsrsetup}). In the electron-target section (ETS)~\cite{sprenger04}, optionally equipped with a thermionic- or photocathode, the electrons are tuned to well defined energies $E_d$ independently of the electron cooler, where an electron beam of constant acceleration voltage precisely defines the ion beam energy. The electron cooler of the TSR yields an electron beam with a density of $1.6\times 10^7$ cm$^{-3}$ and thermal electron energies of $kT_{\perp}\sim10.0$\,meV and $kT_{\|}\sim0.1$\,meV transverse and parallel to the beam direction, respectively. In the ETS the energy resolution of the detuning energy $E_d$ in the co-moving reference frame is mainly limited by the thermal electron velocities. The GaAs photocathode in the ETS provides an electron beam with thermal energies of $kT_{\perp}\sim0.5$\,meV and $kT_{\|}\sim0.03$\,meV~\cite{orlov05}. In the experiments conducted with the photocathode two different electron densities are used, \mbox{$n_{e,l}\,=\,0.63\cdot10^6\,cm^{-3}$} and \mbox{$n_{e,h}\,=\,1.27\cdot10^6\,cm^{-3}$}, which are referred to as the low and high electron density throughout the paper. For some measurements the thermionic cathode is used in the ETS, yielding $kT_{\perp}\sim2.0$\,meV and $kT_{\|}\sim0.045$\,meV at $n_e=2.79\times 10^6$~cm$^{-3}$. 

Neutral fragments stemming from DR events are recorded downstream of the ETS either by a counting or imaging detector (Fig.~\ref{fig:tsrsetup}). For high resolution fragment imaging a detector~\cite{nevo:07} has been placed at \mbox{$\bar{s}=12.24$ m} from the center of the merging region of the ETS, where fragments impinging on a 80-mm diameter microchannel-plate (MCP) at a rate of $<\,$1~kHz produce light spots on the attached phosphor screen. A CCD camera records the light spots at a frame rate of up to 30~s$^{-1}$. In order to ensure that only fragments stemming from a single DR event are recorded in each camera frame, the phosphor screen is switched off as soon as the first fragment arrives, but leaving enough time for the detection of the second particle. The transverse distance $D$ between the two fragments is then determined from the fragment positions of a double-hit event with a resolution of \mbox{$<\,100\, \mu$m}. For DR rate measurements an energy sensitive surface barrier detector can be placed temporarily $\sim$0.5~m in front of the MCP detector.

After each injection the ion beam is stored for 20~s. During the first 7~s of the total storage time period the electron cooler and the ETS are both operated at velocity-matching values to phase-space cool the ion beam. Taking advantage of the twin-merged beam setup the photocathode in the ETS is then detuned to non-zero electron collision energies for 7--17~s from injection before being reset to cooling conditions in the final 3~s. The electron cooler beam remains at zero relative velocity, thereby constantly defining and stabilizing the ion beam energy. The initial 7~s precooling period ensures common ion beam conditions for the subsequent storage time, i.e. with respect to phase-space cooling and the initial molecular ion state distribution. Reference imaging measurements at the end of the initial 7~s and during the last 3~s are acquired, probing the rotational state distribution of the ions through DR at zero detuning energy. The first rotational probe is used to confirm that the same equilibrium state distribution for all measurements is reached before the ETS energy is detuned, while the final reference measurement confirmed the absence of any significant rotational excitation by electrons at non-zero collision energy. In case of the use of the thermionic cathode the total storage time was 15~s and the ETS could be operated continuously at detuning energy without affecting the ion beam quality.

\subsection{2D fragment imaging}
\begin{figure}[bt]
\centering
\includegraphics[width=14cm,keepaspectratio]{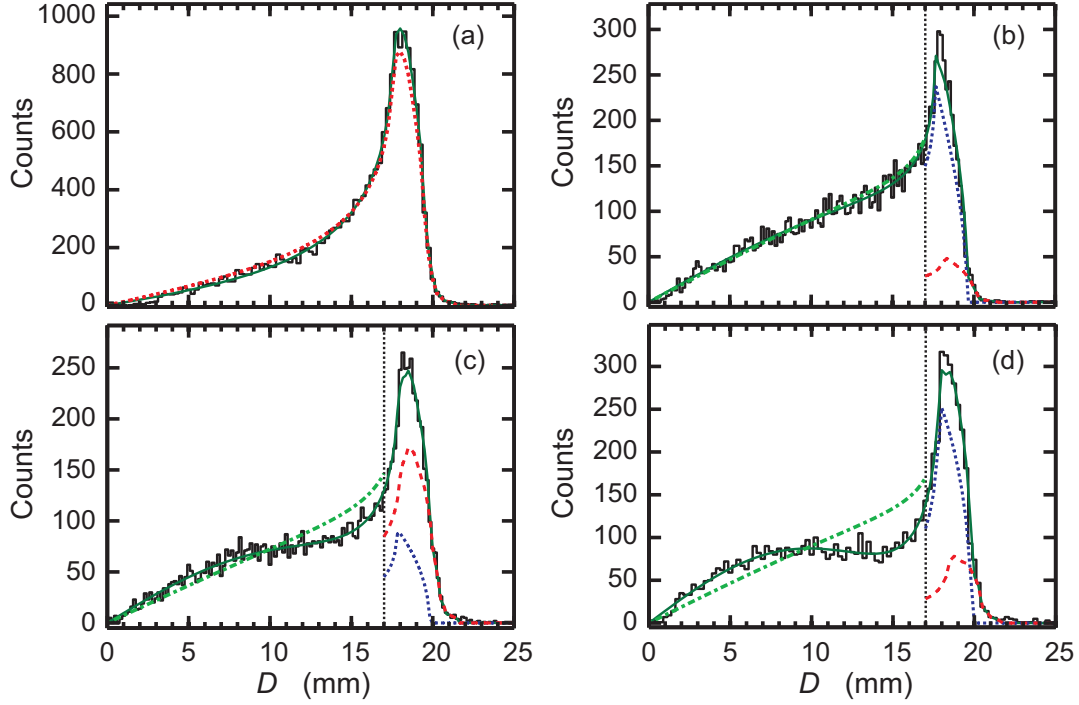}
\caption{\label{fig:2ddist}(color online) Transverse distance distributions
    $F(D)$ for $E_d=0$~meV, thermal cathode (a), $E_d=8$~meV (b), $E_d=25$~meV (c) and $E_d=27$~meV (d), all using the photocathode, with fits applying Eq.(\ref{equ:2ddist}) (solid lines, green). The measured distributions and their best fits are compared in (a) to an isotropic distribution (dotted, red), whereas in (b,c,d) the relative contributions of Legendre polynomials of order $n\leq 2$ (dash-dotted, green) to the total fits are shown for $D<17$~mm, while for $D>17$~mm the relative rotational state contributions of $J\leq1$ (dotted, blue) and $J\geq2$ (long-dashed, red) are displayed.}
\end{figure}

Fragments originating from a DR event arrive after a flight distance $s$ at the detector. Their transverse distance $D$ is then determined through $D=s\delta_J\sin{\theta}$~\cite{amitay:96} by the fragmentation angle $\theta$ relative to the beam direction in the co-moving frame and the maximum laboratory emission angle $\delta_J$. Latter contribution depends on the molecular ion rotational state $J$ through the KER, which is defined by the initial molecular ion and final fragment states as well as the electron collision energy. Averaged over the interaction region in the ETS with $s_1 \le s \le s_2$ ($s_2 - s_1 =L=1.3$~m) two essential properties of the fragmentation process influence the observed transverse fragment distance distribution $F(D)$. Towards the high distance edge, limited by the maximum distance $s_2\delta_J$, the sensitivity of $F(D)$ on the KER increases and is thus dominated by the rotational level contributions. The shape of the total $F(D)$ spectrum at lower distances contains the information on the angular fragmentation pattern. Figure~\ref{fig:2ddist} shows the measured transverse distance distributions at several electron collision energies. Both properties, the rotational state contributions as well as the shape related to the angular distribution, vary distinctly as functions of energy, even for an energy step as small as 2 meV (Fig.~\ref{fig:2ddist}(c) - (d)).

In order to analyze the $F(D)$ spectra quantitatively, the angular distribution $W(\theta)$ of the fragments in the co-moving frame relative to the beam direction is described by Legendre polynomials 
\begin{equation}
W(\theta)\,=\,\sum_n\,a_n P_n(\cos{\theta})
\label{equ:legendre}
\end{equation}
with Legendre coefficients $a_n$ of order n. Since the transverse distance $D$ does not allow to distinguish forward and backward dissociation, i.e. $\theta$ from $\pi - \theta$, only even Legendre polynomials have to be considered and the total distribution is normalized such that $a_0 = 1$.  
For each polynomial $P_n(\cos{\theta})$ and rotational state contribution the transverse distance distribution $F_{J,n}(D)$ is computed by integrating over all event distances $s$. The resulting equations for $n = 0$ and $2$ have been derived earlier~\cite{amitay:96}. For the present analysis it is necessary to consider also the $n = 4$ Legendre polynomial which results in
\begin{equation}
F_{J,4}(D) =\left\{
\begin{array} {l}
{\displaystyle \frac{35\, D}{32\, L\, \delta_J^2}\left( \frac{\Gamma_1^3}{s_1^4}-\frac{\Gamma_2^3}{s_2^4}\right)+\frac{15\, D}{64\, L\, \delta_J^2}\left( \frac{\Gamma_2}{s_2^2}-\frac{\Gamma_1}{s_1^2}\right)+\frac{9}{64\, L \,\delta_J}\left( \arccos{\frac{D}{\delta_J\,s_2}}-\arccos{\frac{D}{\delta_J\,s_1}}\right)}\vspace{0.2cm} \\ 

{\displaystyle \ \hspace{3cm}{\mathrm{for}}\;0\,\le\,D\,\le\delta_J\,s_1} \vspace{0.2cm} \\

{\displaystyle - \frac{35\, D\, \Gamma_2^3}{32\, L\, \delta_J^2\, s_2^4}+\frac{15\, D\, \Gamma_2}{64\, L\, \delta_J^2\, s_2^2}+\frac{9}{64\, L\, \delta_J}\;\arccos{\frac{D}{\delta_J\,s_2}}}\vspace{0.2cm}\\

{\displaystyle \ \hspace{3cm}{ \mathrm{for}}\;\delta_J\,s_1\,\le\,D\,\le\delta_J\,s_2}\vspace{0.2cm} \\ \label{equ:p4}
{\displaystyle \hspace{2cm}{0} \hspace{1cm}{ \mathrm{otherwise}}} \\
\end{array} \right. 
\end{equation} 
with $\Gamma_i$ ($i = 1,2)$ defined as $\sqrt{s_i^2 - (D/\delta_J)^2}$ and $L$ describing the interaction length. The small rotational energy differences allow only to deduce rotationally averaged angular distributions, that is we describe the total, normalized distance distribution $F(D)$ by 
\begin{equation}
F(D)\,=\,\textstyle\sum_J b_J \sum_n a_n F_{J,n}(D)
\label{equ:2ddist}
\end{equation}
with rotational state contributions $b_J$, which are proportional to the population as well as to the DR rate coefficient of the rotational state $J$. 

The parameters $b_J$ and $a_n$ are extracted from a least-squares fit to the data. In a first step, assuming constant, $J$-independent DR rate coefficients, the rotational weighting factors $b_J$ are related to each other through a Boltzman distribution and effective rotational temperatures T are deduced from the fits as a measure of the $J$ states contributing at a given detuning energy $E_d$. In a second step, the parameters $b_J$ are fitted individually with the exception of those for the two lowest rotational levels, $J = 0$ and 1, which are forced in the fit to \mbox{$b_0 = b_1 = b_{01}/2$} as the projected distance distributions for the lowest two rotational states cannot be distinguished due to their small energy difference. This enables us in the further analysis to extract the relative contribution $b_{01}$ of the two lowest rotational levels. In contrast to earlier fragment imaging experiments the fits require Legendre polynomials up to order $n = 4$, while contributions from higher orders were found to be negligible and were therefore set to zero in the final analysis. In the examples shown of Fig.~\ref{fig:2ddist} the best fits are displayed together with some of their individual components. Note that the low-$J$ contribution $b_{01}$ increases by more than a factor of two between 25 meV and 27 meV, i.e. within a small step of only 2 meV. Moreover, the necessity to include the 4$^{th}$ order Legendre polynomial is stressed by comparing the total fit result to the contribution involving only the lowest two Legendre polynomials. 

\section{Fragmentation kinematics} 
\subsection{Rotational state contributions at nearby DR resonances}
\begin{figure}[tb]
\centering
\includegraphics[width=9.5cm,keepaspectratio]{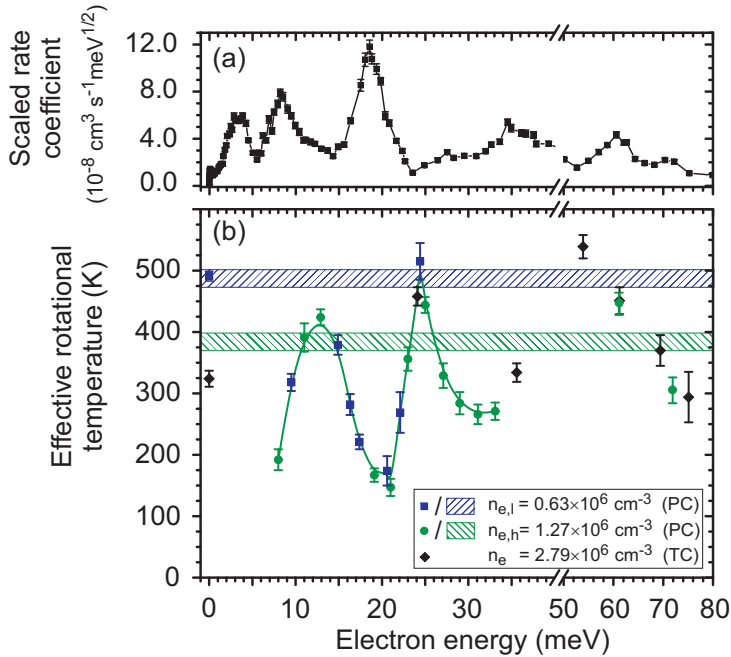}\hspace{1pc}%
\begin{minipage}[b]{5.5cm}\caption{\label{fig:tempvsenerg}(color online) (a) Scaled, rotationally averaged DR rate coefficient $\tilde\alpha(E_d)E_d^{1/2}$ and (b) effective rotational temperature as functions of the detuning energy $E_d$ for the different measurements performed (PC: Photocathode, TC: thermionic cathode). The two bars extending over the full energy range mark the effective rotational temperatures measured at zero detuning energy between 6.5 s and 7 s, i.e. after the initial cooling phase and before jumping to the detuning energy $E_d$.}
\end{minipage}
\end{figure}

In the range below $\sim0.1$~eV the DR cross section of HD$^+$ is believed to be strongly dependent on the initial molecular ion rotational state~\cite{takagi:93,schneider:97} resulting in the pronounced structures observed in the measured rotationally averaged DR rate $\tilde\alpha(E_d)$ (Fig.~\ref{fig:tempvsenerg}(a))~\cite{alkhalili02}. This is convincingly corroborated by the effective rotational temperatures $T$ displayed in Fig. 3(b), which were extracted  from the fragment distance distributions as discussed above. We find that between 7 s and 17 s $T$ varies strongly as a function of energy on a similar narrow scale as $\tilde\alpha(E_d)$ and mostly deviates from the average state contribution $T$ probed by reference measurements at zero detuning energy (shaded bars for the measurement at high and low electron density, respectively). Note, that the effective rotational temperatures deduced at detuning energies $E_d > 8$~ meV do not depend on the different target electron densities employed in the measurements nor on the different initial rotational state distribution reached during the first 7 s of combined electron cooling; this clearly shows that the observed variations of $T$ with $E_d$ are caused by the varying contribution of low and high rotational states to the DR rate. At zero detuning energy, on the other hand, the influence of superelastic collisions (SEC) between rotational states, as reported  in~\cite{shafir07}, is reflected in the electron density dependent equilibrium temperatures obtained after initial cooling. 

To obtain a more detailed view on the rotational states contributing to the DR rate and on the angular fragment distribution we fitted the projected distance distributions with Eq.(\ref{equ:2ddist}) treating $b_{01}$, the individual  $b_J$  for $J > 1$, and the Legendre coefficients $a_2$ and $a_4$ as free parameters (Fig.~\ref{fig:results}). The results show that at certain detuning energies molecular ions in the lowest two rotational states contribute with more than 80\% to the recorded DR events (Fig.~\ref{fig:results}(a)). Furthermore the $b_{01}$ fraction is found to change very rapidly within only a few meV.

\subsection{Anisotropy at non-zero detuning energy}

\begin{figure}[tb]
\centering
\includegraphics[width=9.5cm,keepaspectratio]{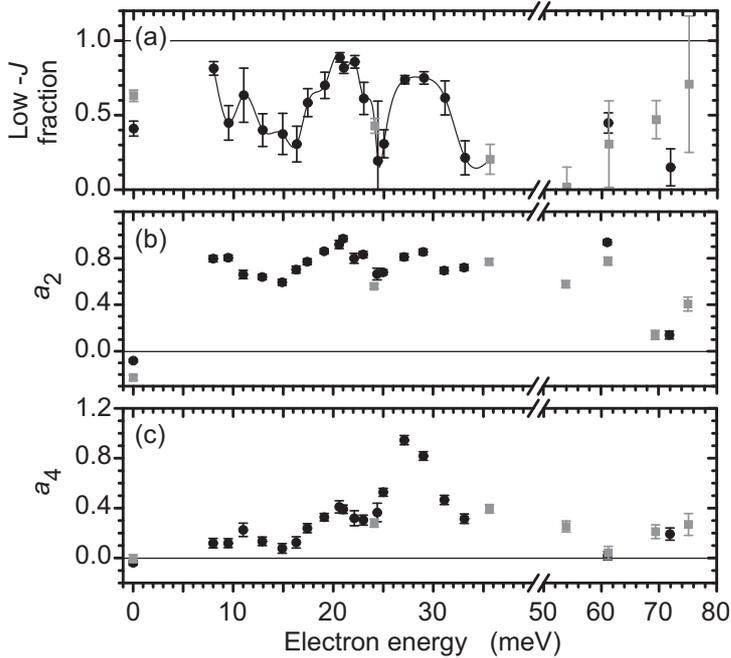}\hspace{1pc}%
\begin{minipage}[b]{5.5cm}\caption{\label{fig:results}  (a) Low-$J$ fraction
    $b_{01}$ (symbols connected to guide the eye), and (b, c) Legendre
    coefficients $a_2$ and $a_4$ as functions of the detuning energy
    $E_d$.  The photocathode is used except for the grey symbols
    indicating use of the thermionic cathode.}
\end{minipage}
\end{figure}

As shown in Fig.~\ref{fig:results}, the Legendre coefficients obtained from the analysis applying Eq.(\ref{equ:2ddist}) yield similar narrow energy dependent structures as $\tilde\alpha(E_d)$, and peaks in the $a_2$ and $a_4$ coefficients generally coincide with enhanced $b_{01}$ factors. The 4$^{th}$ order coefficient $a_4$, observed for the first time in DR angular distributions, turns out to depend particularly sensitive on the collision energy $E_d$ and to decrease with $E_d$ approaching zero eV. The $a_2$ parameter, on the other hand, oscillates slightly around a value of 0.8 for energies above 8 meV. 

Within the axial recoil approximation, which requires the time scale for dissociation to be much smaller than the rotational period, the fragment angular correlation is governed by the partial waves of the incoming electrons leading to dissociation. As the $d$ partial wave is expected to dominate~\cite{giusti:83, novotny:08}, $a_4$ coefficients of the order of $a_4 \approx 2.5$ are expected within this approach independent of the amount of the $s$ wave contribution which is present as well. While the necessity to include Legendre polynomials of up to the order of 4 in the analysis of the projected distance distributions is affirming the importance of $d$ waves in the entrance channel, the $a_4$ coefficients are found to be considerably smaller than expected. In particular one notes, that minima in the measured low-$J$ fraction, i.e. an enhancement of contributions from highly rotating molecules, tend to coincide with minima in the Legendre coefficients, suggesting a loss of anisotropy due to a sizable rotation of the neutralized molecule before dissociation. 
While the direct mechanism of DR is expected to be fast, the indirect DR process~\cite{bardsley68}, involving ro-vibrational excited neutral Rydberg states and thought to be responsible for the pronounced resonance structure in the DR rate coefficient, could well lead to delays in the dissociation comparable to the rotational period, thereby violating the non-rotation assumption of the axial recoil approximation. 

MQDT-calculations of the angular fragment distributions from the DR of HD$^+$ including ro-vibrational couplings are presently being performed~\cite{waffeutamo07}; so far, first results obtained in the axial recoil approximation show reasonable agreement with the data at selected electron energies but generally predict too large anisotropies, indicating again the need to expand the theoretical description beyond the axial recoil limit to arrive at a more detailed understanding of the fragment angular distributions observed in the DR of HD$^+$.

\subsection{Anisotropy at zero detuning energy}

For a spherical symmetric energy distribution around the detuning energy $E_d$ the fragment angular distribution has to be isotropic for symmetry reasons if the detuning energy is adjusted to $E_d = 0$. However, due to the flattened velocity distribution of the electron beam, the spherical symmetry is violated and the fragment distribution may be anisotropic despite of $E_d$ being zero.  

This effect could be observed for the first time in our present measurements (Fig.~\ref{fig:2ddist}(a)). As expected, the anisotropy is found to depend on the transversal electron beam temperature as $T_{\perp} >> T_{\|}$; while the data collected with an electron beam from the photocathode, characterized by $kT_{\perp} \approx 0.5$ meV is still compatible with an isotropic distribution, a small but statistically significant negative $a_2$ coefficient was observed when using an electron beam produced from the thermionic source, which exhibited a perpendicular temperature of $kT_{\perp} \approx 2.0$ meV. Note also that the negative sign of the extracted $a_2$ coefficient is consistent with the results obtained at non-zero detuning energies considering the change from a predominantly longitudinal electron approach to a case where the electrons are approaching mainly from the transverse direction.

\section{Conclusion}
Fragment imaging combined with a twin-merged beam setup opens up a new observational window on the kinematic nature of the DR process. High resolution and energy controlled measurements on a narrow grid of energies enable the study of the DR product kinematics down to a few meV.

For HD$^+$ this technique made it possible to focus on different aspects of the ro-vibrational capture resonances in the region of up to $\sim$~75 meV electron collision energy. Firstly, the KER allowed us to determine the contribution of the lowest two rotational states of the HD$^+$ ion in the recorded DR events. As a function of energy the $b_{01}$ fraction varies strongly, similar to the structure of the rotationally averaged DR rate coefficient $\tilde\alpha(E_d)$. Since the probe measurements confirm a common initial rotational state distribution and exclude different ion beam conditions, the collision energy dependent variations in the low-$J$ fraction thus reflect the sensitivity of the HD$^+$ DR process on the initial ro-vibrational state. 
Secondly, the measured fragment angular distribution reflects within the axial recoil approximations directly the dependence of the electron capture on the orientation of the molecular axis with respect to the incoming electrons. The angular distributions observed here require for the first time 4$^{th}$ order Legendre contributions for a complete description expected if $d$ waves are contributing in the entrance channel leading to DR. However, the $a_4$ coefficients are considerably smaller than expected and vary together with the $a_2$ coefficients on a similar energy scale as the DR rate coefficient $\tilde\alpha(E_d)$. The partial correlation of enhanced contributions of highly rotating molecules and reduced Legendre coefficients might indicate the break-down of the axial recoil approximation.

At zero detuning energy the flattened velocity distribution of the electron beam on the fragment distribution was studied. With an electron beam produced by the thermionic cathode of the electron target, which generates electrons at a higher transverse temperature as the photocathode source, a slight anisotropic fragment angular distribution could indeed be observed as expected in the presence of an asymmetric electron velocity distribution.
\vspace{0.5cm}
\\
{\it Acknowledgments}\\
D. Schwalm acknowledges support by the Weizmann Institute through the Joseph Meyerhoff program. This work has been funded in part by the German Israeli Foundation for Scientific Research (GIF) under Contract No.I-707-55.7/2001.

\end{document}